\documentclass{article}
\usepackage[T1]{fontenc} 
\usepackage[utf8]{inputenc} 
\usepackage{ismir,amsmath,cite,url}
\usepackage{graphicx}
\usepackage{color}
\usepackage{booktabs}
\usepackage{svg}
\usepackage{tabularx}
\usepackage{tabularray}
\usepackage{changepage,threeparttable} 
\usepackage{musicography}
\usepackage{makecell}

\title{FiloBass: A Dataset and Corpus Based Study of Jazz Basslines}

\twoauthors
  {Xavier Riley} {Queen Mary University of London \\ {\tt j.x.riley@qmul.ac.uk}}
  {Simon Dixon} {Queen Mary University of London \\ Centre for Digital Music}

\def\authorname{X.\ Riley and S.\ Dixon}

\begin{document}

\maketitle
\begin{abstract}
We present FiloBass: a novel corpus of music scores and annotations which focuses on the important but often overlooked role of the double bass in jazz accompaniment. Inspired by recent work that sheds light on the role of the soloist, we offer a collection of 48 manually verified transcriptions of professional jazz bassists, comprising over 50,000 note events, which are based on the backing tracks used in the FiloSax dataset. For each recording we provide audio stems, scores, performance-aligned MIDI and associated metadata for beats, downbeats, chord symbols and markers for musical form.

We then use FiloBass to enrich our understanding of jazz bass lines, by conducting a corpus-based musical analysis with a contrastive study of existing instructional methods. Together with the original FiloSax dataset, our work represents a significant step toward a fully annotated performance dataset for a jazz quartet setting. By illuminating the critical role of the bass in jazz, this work contributes to a more nuanced and comprehensive understanding of the genre.
\end{abstract}

\begin{figure*}
    \centering
    \includegraphics[scale=1,width=\textwidth]{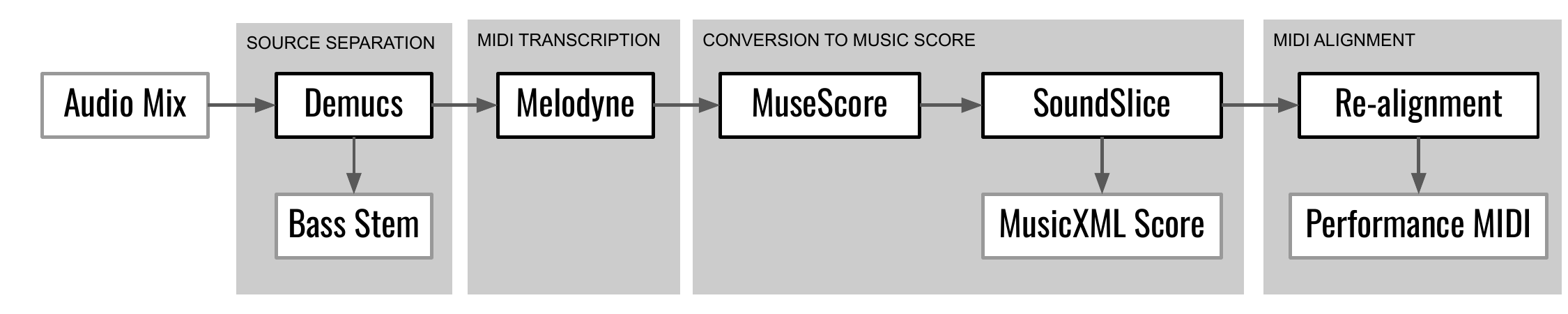}
    \caption{Flow diagram describing the main stages of the proposed method.}
    \label{fig:flow_diagram}
\end{figure*}

\section{Introduction}\label{sec:introduction}

The role of the double bass (also known as the string bass or upright bass) in jazz is nearly ubiquitous as a time keeper, outliner of harmony and as an occasional soloist. A key function is to play ``walking bass'', where the harmony of the song is outlined by playing chord tones on strong beats and linking them with arpeggio, scale or chromatic movements on the remaining beats in the bar. This style has emerged as a way to provide a rhythmic and harmonic foundation to support a soloist. We believe that the harmonic techniques that performers use to outline chord changes could provide important information for enhanced understanding of jazz from an MIR perspective, e.g.\ for generative models. Due to the relatively simple rhythmic vocabulary, this style lends itself to algorithmic approaches which reduce the problem to beatwise pitch predictions, as discussed in Section \ref{sec:related}. However, we recognize that this is a simplified view of bass performance, as bass lines also contain rhythmic subtleties and other nuances which serve to increase the interest and texture of the music over time.

The FiloSax dataset\cite{foster_filosax_2021} addressed a need for high quality annotations\cite{gardner_mt3_2022} to enable downstream tasks like automatic music transcription, score layout and performance analysis. Building on this, we address the need for similarly high quality data relating to the double bass as used in jazz, by turning our attention to the backing tracks used to create that dataset. The backing tracks are taken from the Aebersold series\footnote{\url{http://jazzbooks.com/jazz/JBIO}} and include performances by professional musicians.

Given the high quality of the bass playing on these tracks, we provide fine-grained annotations to allow for detailed stylistic and harmonic analysis.  We believe that this represents the first large scale dataset to include detailed performance timing for jazz bass, which in turn should allow for more realistic generative modelling applications and better results for automatic transcription models.
The transcriptions have been carried out using a semi-automatic pipeline which we describe in Section \ref{sec:methodology}. Each note was checked manually and additionally proof-read by a professional jazz bassist. We also publish the extracted audio stems together with the transcriptions using the SoundSlice platform\footnote{\url{https://www.soundslice.com/}} to allow for easy browsing and evaluation\footnote{\url{https://aim-qmul.github.io/FiloBass/}}. Audio, MIDI and MusicXML artefacts along with the code to produce our analysis are available to download via the same site.

\section{Related work}
\label{sec:related}

Despite the important role of bass in the jazz genre, study of this subject has often relied on fully manual transcriptions which are extremely labour intensive to produce (see \cite{pinheiro_jazz_2018} for an example). To address the need for data on a larger scale, important work was led by Abeßer et al.\ into automatic transcription of bass lines in a jazz context\cite{abeser_deep_2017, abeser_improving_2018, abeser_jazz_2021}. One of the motivations for their work was the idea that accurate bass transcriptions may be used to derive information about the harmony of a song, which in turn could aid with the task of automatic chord estimation. This resulted in 41 automatic bass transcriptions (with manual verification) as part of the Weimar Jazz Dataset\cite{frieler_two_2018} (WJD). These are beat-wise pitch transcriptions, meaning that they are only a partial annotation of the performance, omitting information about rhythmic details, which may limit the use of this dataset in some downstream tasks such as performance analysis or generative modelling. Recent releases of the WJD dataset have included a further 415 fully automatic transcriptions of the bass notes for each beat.

The RWC-Jazz database\cite{goto2004development} (a subset of the widely cited RWC dataset) provides audio and aligned MIDI annotations for 5 pieces, which have multiple recordings across a number of different instrument groupings which include bass. Bass is included on 37 recorded tracks which total around 3 hours of audio, however the audio is synthesised from samples of isolated notes and is mixed rather than provided as individual audio stems. This allows for accurate alignment at the expense of some realism in terms of articulation and dynamic range.

Formal research into walking bass has also focused on rule-based generation for modelling bass performances\cite{shiga_generating_2021}. By incorporating rules described in instructional materials for learning jazz bass, the authors were able to construct a hidden Markov model (HMM) which produced musically relevant results according to subjective listening tests. The authors mention a lack of training data for this task and also note that they were unable to model anything beyond beat-wise pitch estimation.

Outside of the jazz genre, Araz \cite{araz_automatic_nodate} describes a pipeline for transcribing bass lines from electronic music. This approach relies on source separation to extract a bass stem before transcribing it to quantised MIDI. This approach assumes that the music is recorded at a fixed tempo, which is usually the case for electronic genres however this is not usually the case for jazz performances. The MedleyDB\cite{bittner_medleydb_2014} dataset provides a large corpus of multitrack audio recordings. Of these, 71 have been and annotated and resynthesised using the process described in \cite{salamon_analysissynthesis_2017} to produce the MDB-bass-synth dataset. This dataset is primarily aimed at training and evaluating framewise pitch estimation (f0) methods. We also note the IDMT-SMT-Bass dataset\cite{abeser_feature-based_2010} which provides individual recordings of each note on an electric bass with a variety of playing techniques. This may be a good basis for a synthetic dataset to approach similar tasks. A summary of the available datasets is shown in Table \ref{tab:dataset_comparison}.

\begin{table*}[t]
\centering
\resizebox{\textwidth}{!}{
\begin{tabular}{lrrrrrrrrr}\toprule
Name &Annotation Method &Audio sources &Sync. level &Track count &Duration (s) &Note count &Additional Metadata &Scores \\
\midrule
WJD Bass &Automated + Manual &Audio mix &Beat &41 &1851 &5000 &Downbeat, Chord &No \\
WJD v2.2 &Automated &Audio mix &Beat &456 &49010 &122540 &Downbeat, Chord &No \\
MDB-bass-synth &Automated &Audio mix, Audio stems &Frame &71 & 14393 &N/A &None &No \\
RWC-Jazz &Manual &Audio mix &Note &37 &10878 &19183 &Downbeat, Chord &No \\
IDMT-SMT-Bass &N/A &Individual notes &N/A & &12960 &4300 &None &No \\
FiloBass (ours) &Automated + Manual &Audio mix, Bass stem &Note &48 &17880 &53646 &Downbeat, Chord &Yes \\
\bottomrule
\end{tabular}
}
\caption{Comparison of existing bass datasets}
\label{tab:dataset_comparison}
\end{table*}

\section{Methodology}
\label{sec:methodology}

We now describe the process used to create the dataset which is summarised in Figure \ref{fig:flow_diagram}. We would like to emphasise that the work was carried out by the main author, a semi-professional bassist, and later checked and verified by another professional jazz bassist. Despite the use of automatic methods, every note was checked manually at least twice as a result. While this process was expensive in terms of time spent, the resulting increase in accuracy will provide a solid foundation for future methods.

\subsection{Audio Recordings}

All of the 48 backing tracks in this dataset are recorded in a standard format using professional jazz musicians. Details of the performers are shown in Table \ref{tab:performers}. They feature a jazz trio (piano, bass and drums) with bass panned to the left, drums panned centrally and piano panned to the right. This allows for convenient separation of bass and drums by using a single channel of audio. We are able to further isolate this single channel to obtain a bass stem using the Demucs source separation tool\cite{rouard_hybrid_2022}. The producers of these tracks (Aebersold) have a catalogue of over 1300 tracks recorded in a similar fashion, which means that this approach could be applied to additional tracks in the future.

\begin{table}[t]
\centering
\resizebox{\columnwidth}{!}{
\begin{tabular}{lrrrr}\toprule
Name &Track count &Note count &Born \\
\midrule

Christian Doky     &1~~~~~~ & 1401~~ &1969 \\
Dennis Irwin       &1~~~~~~ & 1321~~ &1951 \\
John Goldsby       &3~~~~~~ & 2564~~ &1958 \\
Lynn Seaton        &1~~~~~~ & 1278~~ &1957 \\
Michael Moore      &1~~~~~~ & 753~~ &1945 \\
Ray Drummond       &2~~~~~~ & 2181~~ &1946 \\
Ron Carter        &5~~~~~~ & 5885~~ &1937 \\
Rufus Reid        &14~~~~~~ & 15280~~ &1944 \\
Steve Gilmore     &10~~~~~~ & 12323~~ &1943 \\
Todd Coolman       &3~~~~~~ & 3952~~ &1954 \\
Tyrone Wheeler    &6~~~~~~ & 5474~~ &~1965 \\
Wayne Dockery      &1~~~~~~ & 1050~~ &1941 \\

\bottomrule
\end{tabular}
}
\caption{Details for each bassist in the dataset}
\label{tab:performers}
\end{table}

\subsection{Transcription}

For the initial transcription of performance MIDI, we opted to use the commercial program Melodyne\footnote{\url{https://www.celemony.com/en/melodyne/what-is-melodyne}}, specifically their ``Melodic'' detection algorithm. This is more typically used for editing vocal performances, however the pitch tracking and note segmentation proved to be broadly accurate for the separated bass stems. The program also offers a convenient interface to edit onsets and pitches manually in cases where the automatic analysis was judged to be incorrect. Each of the 48 scores were loaded into Melodyne and manually corrected where necessary.

To produce a score from the performance MIDI we employed a multi-step process. The first step was to import the existing downbeat annotations from the FiloSax dataset into Melodyne. We then used the ``Make tempo constant'' feature of Melodyne to produce a new file in which variations in the tempo were removed and the note positions rescaled accordingly. For those without access to Melodyne, we note that a similar result could be achieved using the \texttt{adjust\_times} function from the PrettyMIDI library\cite{raffel_intuitive_nodate}.

From this constant tempo version, we export a MIDI file from Melodyne and then import this into MuseScore 3\footnote{\url{https://musescore.org/en}} using their MIDI import procedure. This was found to work better when the tempo was made constant first. This yields a score representation, however the variations in timing can produce non-idiomatic representations in the score which need to be corrected. This was done by exporting MusicXML and performing the final corrections using the SoundSlice platform, which allowed the transcription to be edited with reference to the synchronized audio from the original bass stem. Chord annotations are then copied from the FiloSax metadata and all 48 scores were checked by a professional jazz bassist to ensure accuracy and readability.

Finally, we used the alignment method proposed by Nakamura et al.~\cite{nakamura_performance_nodate} to realign the final score representation to the original MIDI performance data. This step is necessary to obtain a 1-to-1 correspondence in note annotations between score and performance MIDI. However, after working with these annotations we found that the timing information in the performance MIDI produced by Melodyne was not of sufficiently high quality. This resulted in issues when evaluating automatic transcription methods (see \ref{sec:amt}). To improve the alignment quality further, we align the MIDI to the model activations of a pre-trained guitar transcription model following the work of Maman and Bermano\cite{maman_synchronizing}. The realigned MIDI outputs are included in the final dataset.

\subsection{Repeated Passages}

During the construction of the original FiloSax dataset, one of the objectives was to capture a consistent amount of saxophone data for each track. Since the original backing tracks varied in length, the authors edited the original backing tracks to repeat certain sections (usually complete choruses) in order to meet their criteria. This impacts the production of this dataset in that some passages are repeated exactly, however they were transcribed by treating them as a complete performance. This may lead to slight variations in how the rhythmic figures are notated which may be an issue for certain downstream tasks, for example introducing a bias in generative models. We recognise this and will provide instructions on how to remove the repeated sections if desired. Otherwise we provide transcriptions for each track in its entirety to allow for easy alignment with the existing FiloSax data.

\subsection{Double Stops, Grace Notes and Ghost Notes}

The source material used for this dataset is predominantly monophonic in nature, however the performers do make use of double stops (polyphony) in some places. We have transcribed these in the score and alignments but we also provide a monophonic version of the dataset with a view to ease of use in downstream tasks.
The use of effects such as grace notes (extremely short notes) or ghost notes (where the string is partially or fully dampened to produce a percussive sound) is prevalent throughout the dataset and these can be viewed as an important aspect of the style. A guiding principle for producing the score representation is that they are readable by a sufficiently experienced bassist. With this in mind, we have notated ghost notes where these can be clearly heard on the recording however in cases where these effects were judged to be subtle or fleeting we have omitted them. We understand that this approach could be seen as subjective but we did so to prioritise the goal of making a readable and idiomatic score output over a completely consistent yet less readable score.

\subsection{``Common Practice'' versus Real Performance}

The backing tracks used to create this dataset were originally conceived as practice aids for instrumental soloists. As such, the performances on these tracks could be viewed as a sort of ``common practice'' of jazz accompaniment. The performers focus on outlining chord changes and rhythms clearly to allow the soloist to focus on their role. This aspect of the data makes it a valuable example for studying how these accompaniments are constructed. However, they may not be entirely representative of performances from live or studio recordings, as musicians may be more inclined to take musical risks in those settings. For this reason, the figures that we derive in our later analysis might not be fully representative of live or studio performance. A comparison is a potential area for exploration in future work.

\subsection{Dataset Contents and Distribution}

The final dataset comprises 48 tracks with contents as follows: Melodyne project files, audio mixes, isolated bass stems (from source separation software), performance-aligned MIDI with velocity information, and music scores in MusicXML format. We also include metadata which was compiled as part of the FiloSax dataset which includes timings for chords, sections, beats and downbeats.

As discussed in \cite{foster_filosax_2021}, the backing tracks themselves are subject to copyright restrictions so we are unable to release these. However, we provide instructions on how to obtain the files from the original provider. All other assets (including the source-separated stems) will be made freely available to researchers.

\section{Analysis}

We now present a corpus analysis of the data in which we demonstrate the potential for insights on a musical level. As a starting point, we seek to answer some queries about the harmonic and rhythmic functions of a typical walking bass line as represented in the data. A number of commercial jazz bass methods from different authors are summarised in \cite{pinheiro_jazz_2018} which we will refer to where appropriate. All analyses which follow were derived from the dataset by converting note-level information to a Pandas\cite{mckinney2010data} dataframe using the Music21 Python library\cite{music21}. The queries used to perform the analysis will be released alongside the dataset.

\subsection{Chord Degrees Used in Bass Line Construction}
\label{ssec:chord_degrees}

As jazz performance is a cultural practice, a strict set of rules for bass line construction has not been established. However, given the size of the proposed dataset we can start to provide a quantitative analysis of the choices made by performers during their improvisations.

\begin{figure}

\begin{minipage}[t]{1.0\linewidth}
  \centering
  \centerline{\includegraphics[width=\textwidth,trim=0 10 0 10]{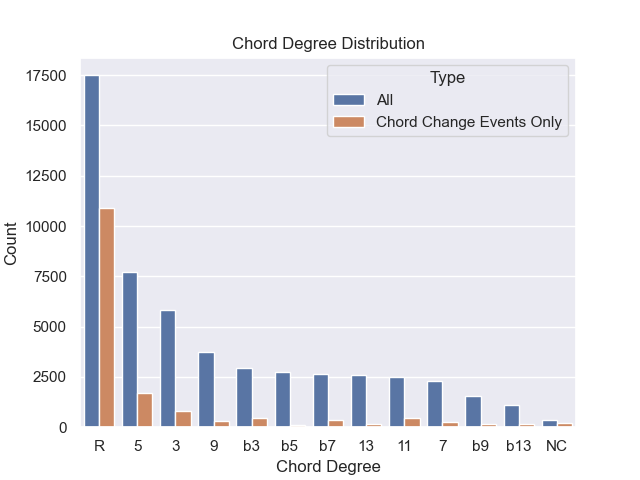}}
  \begin{center}
  \caption{Global distribution of chord degrees\label{note-distribution-global}}
  \end{center}
\end{minipage}

\end{figure}

Concerning the question of which chord degrees are favoured by the player, we analyse the function of each note in the dataset as it relates to the chord being played underneath it. In Figure \ref{note-distribution-global} we see that bassists will favour the root note of the chord when constructing walking bass lines, as these are used in 32.7\% of all notes played. This is rather basic from a musical perspective, but we can now point to data that bolsters existing empirical observations. When we examine the note played at each new chord change event, we see from Figure \ref{note-distribution-global} that the use of chord roots is even more prevalent, with the proportion rising to 67.9\% of the total. This reflects the role of the bass in outlining the harmony of the song.

\subsection{Use of Rhythmic Fills versus Quarter Note Pulse}

In his educational method book, bassist Ron Carter \cite{carter_building_nodate} describes the process of adding rhythmic interest, or ``fills'', to a line. However, he cautions the student ``not to overdo'' their use before advising that: ``personal tastes and judgement will govern this area of your playing''. We can make an attempt to quantify this more precisely by examining what percentage of measures in the dataset contain a simple set of 4 quarter notes, and which deviate from this. We find that 62.81\% of measures are indeed 4 quarter notes. While this is not a substitute for developing good taste, knowing this percentage might help in guiding a more analytical player.



\subsection{Deriving Common Patterns}

\begin{table}[ht]
\begin{tblr}{
      colspec={Q[m]Q[m]Q[m]}
    }
    \hline
    \thead{Pattern} & \thead{Count} & \thead{\% of total} \\
    \hline
    \raisebox{-.5\height}{\includegraphics[clip, trim={8.44cm 23.7cm 7.42cm 4cm}]{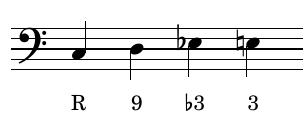}} & 360 & 4.6\% \\
    \raisebox{-.5\height}{\includegraphics[clip, trim={14.33cm 23.7cm 1.53cm 4cm}]{figs/FiloBass-Top-n-grams.pdf}} & 195 & 2.5\% \\
    \raisebox{-.5\height}{\includegraphics[clip, trim={2.13cm 19cm 13.73cm 8.8cm}]{figs/FiloBass-Top-n-grams.pdf}} & 113 & 1.43\% \\
    \raisebox{-.5\height}{\includegraphics[clip, trim={8.44cm 19cm 7.42cm 8.8cm}]{figs/FiloBass-Top-n-grams.pdf}} & 113 & 1.43\% \\
    \raisebox{-.5\height}{\includegraphics[clip, trim={14.33cm 19cm 1.53cm 8.8cm}]{figs/FiloBass-Top-n-grams.pdf}} & 111 & 1.41\% \\
\end{tblr}
\caption{The five most common chord degree n-grams for 7898 chord instances lasting 4 or more beats. Examples are notated in C major for illustration.}
\label{tab:degreetable}
\end{table}

The annotations in this dataset also allow us to examine sequences of chord degrees that are commonly used in bass line construction. Over the 6400 chord symbols annotated, 3900 distinct patterns of chord degrees over chords are played. The 5 most common patterns for a chord lasting 4 beats are shown in Table \ref{tab:degreetable}. From these we can see a preference towards using tones from major and minor triads (i.e.\ 1, $\flat$3, 3 and 5). Given that the root movements in jazz are often perfect 4ths apart, we see that a number of the patterns approach the 4th via tones or semitones (i.e.\ from $\flat$3, 3, $\flat$5 or 5). This analysis of patterns only considers the chord degree, however a more detailed examination of patterns including sequential ideas and motifs is a subject of future work.

\subsection{Semitone and V-to-I Approaches}

In ``Creating Jazz Basslines'', author Jim Stinnet emphasises the use of semitone approaches. This is where target notes which fall on strong beats or chord changes are preceded by a note which is a semitone above or below the target (described in \cite{pinheiro_jazz_2018}). In this dataset we observe that this is indeed common, with ascending and descending semitones being the most often used intervals overall as shown in Figure \ref{interval-distribution}. For notes which land on chord changes, semitone approaches are even more prevalent. We summarise the most common intervals to approach chord changes (target tones) in Table \ref{tab:approach_tones}.

In the ``Walking Bassics'' by Fuqua, Zisman, and Sher (described in \cite{pinheiro_jazz_2018}), the authors advocate the use of V to I movements for students however our data suggests that this is relatively uncommon in practice (9.66\% ascending a perfect fourth and 4.30\% descending). This is an interesting example of an idea that seems intuitive in theory (V to I is a strong bass movement for walking bass) but is not reflected in practice.

\begin{table}[ht]
    \centering
    \begingroup
    \renewcommand{\arraystretch}{1.5}
    \begin{tabularx}{8.2cm}{ 
   >{\raggedright\arraybackslash}X 
   >{\raggedleft\arraybackslash}X 
   >{\raggedleft\arraybackslash}X 
   >{\raggedleft\arraybackslash}X }
        \toprule
        \thead[l]{Approach} & \thead[l]{Interval to target} & \thead[r]{Count} & \thead[r]{\% of total} \\
        \midrule
        D\musFlat{} $\searrow$ {C} & $-1$ & 4318 & 26.75 \\
        B\musNatural{} $\nearrow$ {C} & $+1$ & 3384 & 20.97 \\
        B\musFlat{} $\nearrow$ {C} & $+2$ & 1921 & 11.90 \\
        G $\nearrow$ {C} & $+5$ & 1560 & 9.66 \\
        C $\rightarrow$ {C} & $0$ & 1172 & 7.26 \\
        G $\searrow$ {C} & $-7$ & 694 & 4.30 \\
        D $\searrow$ {C} & $-2$ & 656 & 4.06 \\
    \end{tabularx}
    \endgroup
    
    \caption{The most common intervals used to approach a chord change (totalling 16141 events). For illustration all approaches are shown relative to a target tone of C.}
    \label{tab:approach_tones}
\end{table}

\subsection{Step, Leap or Staying Put?}

\begin{figure}

\begin{minipage}[t]{1.0\linewidth}
  \centering
  \centerline{\includegraphics[width=\textwidth]{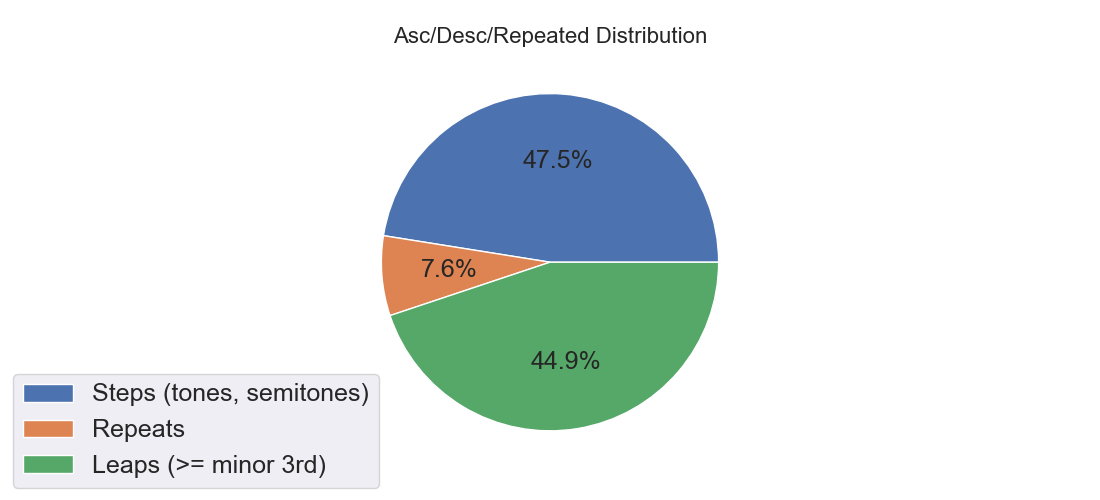}}
  \begin{center}
  \centerline{\includegraphics[width=\textwidth,trim=0 10 0 10]{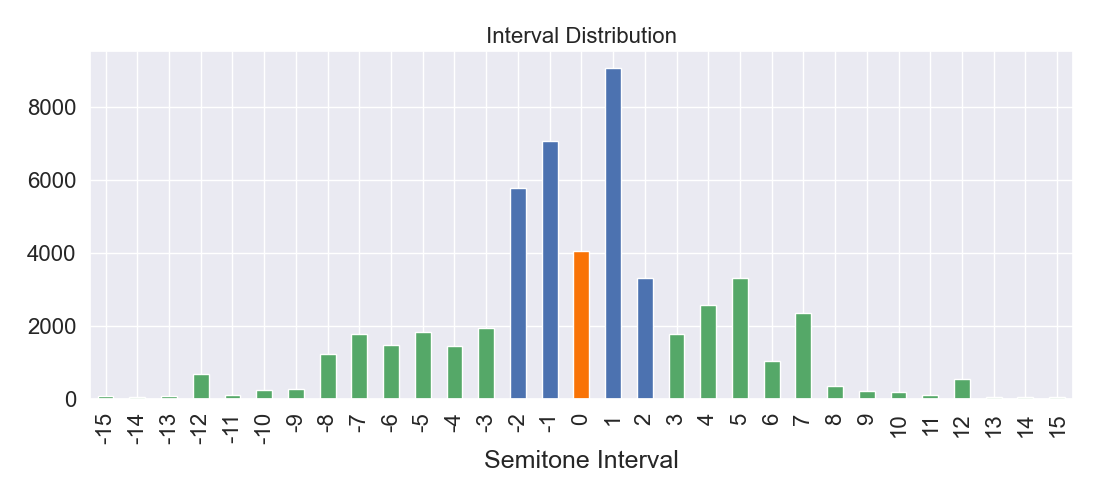}}
  \caption{Distribution of intervals, grouped as step-wise movements (2 semitones or less), leaps (3 or more semitones) or repeats (no change from the preceding note).}
  \label{interval-distribution}
  \end{center}
\end{minipage}

\end{figure}

As we have seen in Section \ref{ssec:chord_degrees}, in the majority of cases the performer will aim to play root notes when a new chord arrives but this leaves the question of how these root notes are typically connected together into a musically pleasing line. From the data, we can examine whether performers tend to use step-wise motion (tones and semitones), larger intervalic leaps (minor thirds or greater) or whether they choose to repeat a note. Looking at Figure \ref{interval-distribution} we see that there is a slight preference toward using step-wise motion (the largest group at 47.5\%). Viewing the interval distribution plot we can also see that intervalic leaps are slightly more likely when the line is ascending especially for the interval of 5 semitones which corresponds to a perfect fourth.

\subsection{Melodic Contour}
\label{ssec:changing_direction}

\begin{figure}[ht]

\begin{minipage}[t]{1.0\linewidth}
  \centering
  \centerline{\includegraphics[width=\textwidth,trim=0 10 0 10]{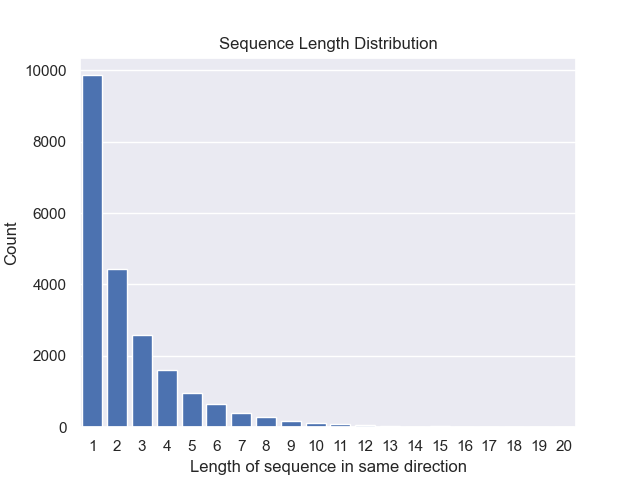}}
  \begin{center}
  \caption{Sequence length (number of intervals) of lines maintaining a constant direction.}
  \label{directions}
  \end{center}
\end{minipage}

\end{figure}

The performer has a number of parameters available when improvising a bass line, one of which is the direction of the line. Sigi Busch (summarised in \cite{pinheiro_jazz_2018}) refers to the idea of ``voice leading'' within a bass line to link important chord tones while maintaining a direction, but none of the other methods summarised in \cite{pinheiro_jazz_2018} advise on how to choose directions or when to change them. Referring to the data now, we can see in Figure \ref{directions} that a high number of changes in direction is preferred, with the mean length of a sequence before a change falling at 2.46 notes. Intriguingly, the distribution of sequence lengths exhibits a power law. This phenomenon has been observed in several cases when analysing symbolic music corpora\cite{rafailidis_power_2008} but to our knowledge this is the first evidence in relation to walking bass lines.

\section{Automatic Transcription Baseline}\label{sec:amt}

Using the accurate alignment data we have collected, we provide initial results for automatic note transcription --- a bass line baseline. An exhaustive appraisal of transcription accuracy is beyond the scope of this work but we hope these results will encourage the use of this dataset in related future work.

We use the \texttt{mir\_eval}~\cite{mireval} library to calculate precision, recall, F-measure and overlap scores. A default threshold of 50ms was used and only onset timings were considered. This is due to the difficulty of assessing offsets, as described in \cite{bittner_basic_pitch}.
Three methods are examined for this task; the ``Basic Pitch'' package described in \cite{bittner_basic_pitch}, the ``CREPE Notes'' method proposed in \cite{crepe_notes} and the commercial software Melodyne using the ``Melodic'' algorithm. The results from Melodyne were not manually corrected for this evaluation. Results for all methods are shown in Table \ref{fig:amt-results}.
We see from these results that the proprietary commercial software outperforms the best research solutions for this dataset, however a significant amount of work is required to correct the remaining errors. During this work we also appreciated the Melodyne UI for note editing during our manual correction process. We note that similar projects in future may benefit from open source tools that allow a more streamlined note correction workflow.

\begin{table}
\begin{tabular}{r|l|l|l}
 & CREPE Notes             & Basic Pitch & Melodyne \\
\hline
R$_{no}$               & $74.11\pm12.09$  & $81.28\pm6.26$ & $79.52\pm14.77$ \\
P$_{no}$            & $71.81\pm13.33$  & $51.40\pm6.28$ & $78.48\pm15.41$ \\
F$_{no}$            & $72.89\pm12.68$  & $62.73\pm5.55$ & $78.95\pm15.02$ \\
O              & $78.77\pm2.68$  & $65.24\pm4.51$ & $87.94\pm3.91$ \\

\end{tabular}

\caption{\label{table:amt-results}%
Automatic note transcription results for FiloBass, showing mean scores and standard deviation for Recall, Precision, F-measure and Overlap. Only onsets were evaluated and a timing tolerance of 50ms was used.}
\label{fig:amt-results}

\vspace{-4mm}

\end{table}

\section{Discussion and Future Work}

In collating a dataset and performing a corpus analysis with reference to jazz bass methods, we hope to have provided useful insights into the role of the bass in jazz. The analysis provided here is not exhaustive however, and we hope that future research can reveal more about the mental model that performers use when constructing their bass accompaniment. In particular we hope to examine the role of timing, dynamics and use of sequential ideas in further work. We are also interested in pairing the FiloBass data with the FiloSax data for further analysis. The relationships between bass line and melody in a jazz setting could be explored further, with a view to developing more realistic generative models for both bass lines and solos.

We believe that the dataset has a wide number of potential uses beyond musicological analysis. Recent work on automatic music transcription (AMT) has highlighted that performance can be improved as more data is made available\cite{gardner_mt3_2022} and this dataset can help to address this need.

An additional task which we hope to address in future is that of automatic chord estimation (ACE). Following the hypothesis of Abeßer et al.\ \cite{abeser_deep_2017}, we believe that this data could be used to train a system to estimate chords from the bass line directly. Chord estimation is a particularly challenging task in the jazz setting due to the rich harmonic vocabulary so novel approaches here may be welcome.

The scores which were produced as part of this data should also be valuable to researchers, as they provide a potential source of training data and evaluation for monophonic score processing tasks. In particular, they will be useful for rhythmic parsing (quantisation), automatic score layout and related sub-tasks such as spelling of accidentals.

\section{Conclusions}

We present FiloBass: a new dataset for jazz bass lines. Making use of the detailed annotation data, we are able to demonstrate a quantitative approach to reinforce traditional musicological analysis of the role of the bass in jazz performance.

Through examination of this dataset we demonstrate that a number of rules put forward in jazz bass method books are supported by larger scale data. These can be summarised as follows: the root note of the chord is usually played on the first beat of a new chord; this root is approached via a semitone step where possible; the rhythm comprises a quarter note pulse most of the time; a balance is maintained between ascending and descending contours. We are aware though, that any analytical project of this sort cannot be truly comprehensive and can only offer a guide to the performer. The musical context and the taste and experience of the musician will determine when to follow the ``default'' most likely path and when to choose a different route.

\section{Acknowledgements}

The first author is a research student at the UKRI Centre for Doctoral Training in Artificial Intelligence and Music, supported by UK Research and Innovation [grant number EP/S022694/1].


\bibliography{FiloBass}

\end{document}